# Formation of femtosecond laser induced surface structures on silicon: insights from numerical modeling and single pulse experiments

T. J.-Y. Derrien*[1], R. Torres[1], T. Sarnet[1], M. Sentis[1] and T.E. Itina[2]

[1]Lasers, Plasmas and Photonic Processes Laboratory (LP3), UMR 6182 CNRS – Université de la Méditerranée, Parc Scientifique et Technologique de Luminy, 163 Avenue de luminy - C. 917, 13288 Marseille cedex 9, France

[2]Hubert Curien laboratory (LaHC), UMR 5516 CNRS - Université Jean Monnet, Bat. F, 18 rue du Professeur Benoit Lauras, 42000, Saint Etienne, France

Corresponding author: derrien@lp3.univ-mrs.fr

**Abstract**

Laser induced periodic surface structures (LIPSS) are formed by multiple irradiation of femtosecond laser on a silicon target. In this paper, we focus and discuss the surface plasmon polariton mechanism by an analysis of transient phase-matching conditions in Si on the basis of a single pulse experiment and numerical simulations. Two regimes of ripple formation mechanisms at low number of shots are identified and detailed. Correlation of numerical and experimental results is good.

# Introduction

Femtosecond laser has opened many promising opportunities in texturing semiconductors for photovoltaics [1, 2], microelectronics [3], microfluidics [4] and microbiology [5, 6]. Thus, by using multiple irradiations, it is possible to create conical structures, so-called "Black Silicon" that are famous for their absorption properties in IR and visible range. However, the texturing process reveals several structures along the multiple irradiations [7, 8].

In this paper, we focus on the formation mechanisms of parallel structures named "ripples" which generally appear perpendicular to laser linear polarization after one or several shots, using laser fluence near melting threshold in Si. Different mechanisms have been proposed to explain the formation of such periodic structures. The most well known describes the reaction of a rough surface to the laser field by formation of a periodic pattern of energy deposited on surface [9]. Some insights have been considered for femtosecond interaction case, in which an electron-hole plasma [10] is created due to the transition of electrons from valence to conduction band [11]. For instance, Bonse et al [12] presented a reduction of the period lower than the laser wavelength for free-electron density near plasma critical density, confirming previous signs of plasmonic behavior of matter under femtosecond irradiation [13, 14]. Non-resonant processes have been also proposed in [15] through a modification of surface tension and thermal instabilities coming from nanosecond laser patterned surfaces [16]. A numerical analysis of the different models became unavoidable to elucidate the proper structure formation scenarios.

In this article, we discuss such effects as the surface waves, their properties and the possibility of exciting plasmon resonance. The role of those processes is discussed on the basis of our results for single pulse experiments, and numerical simulations. The experiments present concentric waves on Si with a ripple with a periodicity close to laser wavelength ("Low Spatial Frequency LIPSS" or LSFL [17]). An attempt to distinguish different regimes of laser fluence for very low number of shots is presented for laser fluences near melting and ablation thresholds.

**Calculation results and comparison with experimental findings**

First, we compare the result of a single pulse experiment showing periodic structures and that of a simulation performed by using the classical FDTD (*Finite Difference Time Domain*) approach that solves Maxwell equations [18].

Figure 1(a) shows a SEM picture of the surface resulting from the interaction of a single pulse of 800 nm wavelength, 100 fs of FWHM laser pulse duration, at normal incidence, laser fluence of 1.15 J/cm², with a (100) oriented crystalline polished bulk silicon sample. The precise experimental protocol is described in [19], but the number of shots is controlled using a Pockels cell. The resulting spot shape is top-hat. One can observe the formation of periodical concentric structures, which are more pronounced perpendicularly to the laser linear polarization direction. In addition, the Figure clearly shows that the waves started to propagate from well-defined points, or propagation centers. Those centers are surrounded by two types of concentric periodic structures. Two first oscillations have a period of 838 ± 56 nm. The other oscillations have a period of 775 nm ± 25 nm.

Figure 1b presents the calculation result obtained by using a 3D FDTD method [18]. Here, the absorption of a top-hat laser pulse by bulk silicon is modeled. One can see the effect of the wave scattering by hemispherically shaped holes with radius of 400 nm. As proposed in the reference [20], a subwavelength hole or a slit can act as an electric dipole. It is demonstrated that the profile of the integrated intensity deposited on silicon is periodic as a result of summation of the incident laser pulse and the surface waves. The calculated period is $232 \pm 10$ nm, which corresponds to $\lambda/n$, wavelength of a wave propagating in ground state Silicon. However, the experimentally measured period is close to the laser wavelength. This result was attributed to the surface plasmon polariton by several authors [12, 13, 14]. A surface plasmon polariton is a collective oscillation occurring in the free electron-hole plasma only under very strong conditions that should be better verified. Therefore, in what follows, we check the dielectric index and the conditions for the phase-matching of the resonant process.

It is well-known that surface plasmons can be excited at a metal-insulator interface. Thus, the first condition on the dielectric index of the laser-excited target is that its real part has to be lower than zero as in the case of the metal-dielectric interface [21]. Because of the plasma behavior of the free carriers, it is possible to consider a transient metallic state of the silicon under femtosecond irradiation [10]. The second condition is that the absolute value of the real part of "metal" medium has to be higher than the dielectric index of the "insulator" part of the interface, in order to respect the theoretical limit of light velocity in a medium [22, 23].

Figure 2 shows the period of a SPP as a function of free-electron-density. One can observe that SPP period is strongly sensitive to collision frequency. Periodic structures can be attributed to plasmons if the free-carrier density grows above the critical density, from which the plasma starts to strongly reflect the laser energy. We note that if the Beer-Lambert law is used for femtosecond laser absorption, the free-electron density can be overestimated because one neglects the fact that the transient electron plasma formation affects laser propagation. As a result, the thermal coupling between free carriers and lattice is maximal at the critical density. Therefore, a full Maxwell equation system has to be solved in order to determine if the critical density is a physical limit or can be bypassed during the interaction. Considering this effect, the calculations show that the plasmon model could explain the ripple periods slightly above the laser wavelength, the ones which are observed on

Figure 1a.

To clearly examine conditions of surface plasmon excitation,

Figure 3 defines the minimum free electron density required for the surface plasmon resonance and SPP excitation. By solving the equation $\mathrm{Re}(\varepsilon) = 0$ by using the Drude dielectric function [24], it is possible to extract a solution of free-electron density as a function of collision frequency, for which the metallic state is established (see the green dashed curve). The second condition on real part of dielectric index is $|\mathrm{Re}(\varepsilon)| \geq 1$ [23], which represents the theoretical limit of light velocity. A similar solution is built considering this condition. The half-space above the curves defines the free electron density necessary to excite a surface plasmon polariton. In addition, it is limited by the number of available electrons $Z^* \cdot n_v$ in the valence band, where $Z^*$ is the effective ionization number of per atom.

Figure 4 (a) and (b) show two formulations of the same phase-matching condition.

Figure 4(a) presents the angle of incidence required for phase matching at a given free electron density.

Figure 4(b) shows the free-electron density, for which phase matching occurs at a given incidence angle. The green dashed curve defines the upper limit of free electron density for which total reflection angle is higher than SPP resonance angle. The phase-matching angle has to be lower than the critical angle for which one has total reflection on the surface, derived from Snell-Descartes law $\theta_{cr} = \arcsin\left(\frac{1}{\sqrt{\varepsilon}}\right)$ at vacuum/silicon interface. By solving the equation $\theta_{spp} = \theta_{cr}$, one finds a unique threshold density $n_{cr}^{SPP} = 2.86 \ 10^{28} \mathrm{m}^{-3}$, above which the coupling is not possible [see the green curve in

Figure 4 (b)].

Those results show that there is a possibility to satisfy the phase-matching condition and to excite a surface plasmon polariton at the interface between the air/vacuum and the bulk silicon under a femtosecond irradiation. However, the following points should be underlined

(i) Phase-matching condition requires a small modulation, so that the incidence angle of the laser wave with the modulation is larger than 10°. This result explains why a seed is required for ripples to appear on a silicon sample irradiated by a single femtosecond laser pulse, as it has been reported experimentally in this paper and in literature [12, 25]. A localized surface plasmon can be excited by an interaction with a defect and thus is able to couple with surface plasmon polariton [22].

(ii) Condition of phase-matching imposes a maximum free-electron density of $n_{cr}^{SPP} = 2.86 \; 10^{28}$ m$^{-3}$.

(iii) Conditions on dielectric index revealed that the free-electron density has to be larger than $n_e = 1.7 \; 10^{28}$ m$^{-3}$ at usual collision frequency [24].

(iv) The range of possible free-electron densities authorizing SPP excitation is narrow and can be reached as a transitional state, occurring after the electron avalanche ionization process during laser irradiation.

We finally present an attempt to identify two main different regimes of laser fluence, for which scattering by roughness and surface plasmon theories can explain the periodicity of low N resonant laser-induced periodic ripples in Si.

Figure 5 presents a comparison between our numerical calculations and experimental measurements. Two different surface wave models have been introduced in a 2D two-temperature simulation for Si presented in [26], supposing that experimentally observable periodic modulation is of thermal nature. The thick line represents the period measured in lattice temperature profile as a function of fluence in the case of SPP. One can see that the single pulse experiments quantitatively match the numerical results of surface plasmon polariton theory. Note that the period is higher than laser wavelength for any laser fluence, and slightly decreases above 0.5 J/cm². As shown by the bold curve with error bars, the periodicities provided by the interaction of laser with excited roughness (modeled by the Sipe model [9], and considering the variation of dielectric index as proposed in [12]) are also plotted. Experimental measurements of ripples for low number of pulses (N>1) are shown to qualitatively match the numerical results. Two regimes can be distinguished: for a single laser pulse, we show that a domain of material parameters exists, for which surface plasmon polaritons are excited. In this case, the experimental results correspond to calculated periodicity of surface plasmon. Note, however that a seed is required (e.g. an isolated defect) to respect the phase-matching condition as calculated in this paper. For more than one pulse, the Sipe model applied to the excited Si predicts a periodicity lower than laser wavelength, and the experiments show correlated results as soon as a roughness appears.

## Conclusions

Based on the dielectric index and phase-matching conditions, we have shown that a surface plasmon polariton can be excited on Si during a 100 fs and 800 nm single-pulse irradiation. In particular, the range of the required free-electron density can be reached upon an ultrafast laser irradiation of Si. Such a possibility has been verified by using a full TTM/Maxwell simulation. As a result, the necessary carrier density is high for Si and corresponds to the plasma mirror regime limiting the possible free-electron density.

In addition, the period of a surface plasmon polariton has been calculated depending on laser fluence. We have also shown that the diffraction theory cannot explain the near-wavelength structures in ground state Si.

Two regimes of ripple formation have been defined as follows

(i) For a single laser pulse, the SPP resonance can induce ripples with a period higher than laser wavelength, and a seed is needed to fulfill the phase-matching condition;

(ii) By using several laser pulses, surface roughness is increased and the ripple formation mechanism can be explained by the classical theories, for which ripple periods are lower than laser wavelength.

## Acknowledgements

The authors gratefully acknowledge computer support from the CINES of CNRS, France. David Grojo, Ilya Bogatyrev, and Maxime Lebugle are thanked for fruitful discussions. The experimental assistants Chloé Dorczynski and Maxime Babics are also thanked. We are grateful to Sedao for the critical reading of the manuscript.

## References

[1] Sarnet, T., Halbwax, M., Torres, R., Delaporte, P., Sentis, M., Martinuzzi, S., Vervisch, V., Torregrosa, F., Etienne, H., Roux, L., and Bastide, S. *Proceedings of the SPIE* **6881**, 688119 (2008).

[2] Bonna, K., Sullivan, J., Winkler, M., Sher, M.-J., Marcus, M., Fakra, S., Smith, M., Gradecak, S., Mazur, E., and Buonassisi, T. In *24th European Photovoltaic Solar Energy Conference*, (2009).

[3] Carey, J. *Femtosecond Laser Microstructuring of Silicon for Novel Optoelectronic Devices*. PhD thesis, Harvard University, (2004).

[4] Groenendjik, M. and Meijer, J. *Journal of Laser Applications* **18**, 227 (2006).

[5] Stratakis, E., Ranella, A., Farsari, M., and Fotakis, C. *Progress in Quantum Electronics* **33**(5), 127 – 163 (2009).

[6] Stratakis, E., Ranella, A., and Fotakis, C. *Biomicrofluidics* **5(1)**, 013411 (2011).

[7] Birnbaum, M. *Journal of Applied Physics* **36**(11), 3688 (1965).

[8] Crouch, C., Carey, J., Shen, M., Mazur, E., and Génin, F. *Applied Physics A* **79**, 1635 (2004).

[9] Sipe, J., Young, J. F., Preston, J., and Driel, H. V. *Physical Review B* **27**, 2 (1983).

[10] Sokolowski-Tinten, K., Bialkowski, J., Cavalleri, A., Linde, D. V. D., Oparin, A., ter Vehn, J. M., and Anisimov, S. *Physical Review Letters* **81**, 1 (1998).

[11] Sundaram, S. and Mazur, E. *Nature Materials* **1**, 217 (2002).

[12] Bonse, J., Rosenfeld, A., and Krüger, J. *Journal of Applied Physics* **106**, 104910 (2009).

[13] Miyaji, G. and Miyazaki, K. *Optics Express* **16**, 20 (2008).

[14] Huang, M., Zhao, F., Cheng, Y., Xu, N., and Xu, Z. *ACS Nano* **3**, 12 (2009).

[15] Ben-Yakar, A., Harkin, A., Ashmore, J., Byer, R., and Stone, H. *Journal of Physics D: Applied Physics* **40**, 1447 (2007).

[16] Bäuerle, D. *Laser Processing and Chemistry*. Springer, (2000).

[17] Bonse, J., Munz, M., and Sturm, H. *Journal of Applied Physics* **97**, 013538 (2005).

[18] Lumerical. FDTD Solutions, http://www.lumerical.com.

[19] Torres, R., Vervisch, V., Halbwax, M., Sarnet, T., Delaporte, P., Sentis, M., Ferreira, J., Barakel, D., Bastide, S., Torregrosa, F., Etienne, H., and Roux, L. *Journal of Optoelectronics and Advanced Materials* **12**, 3 (2010).

[20] Lévêque, G., Martin, O., and Weiner, J. *Physical Review B* **76**, 155418 (2007).

[21] Maier, S. A. *Plasmonics, Fundamentals and Applications*. Springer, (2007).

[22] Zayats, A., Smolyaninoy, I., and Maradudin, A. *Physics Reports* **408**, 131–314 (2005).

[23] Raether, H. *Surface plasmons on smooth and rough surfaces and on gratings*. Springer-Verlag, (1986).

[24] Bulgakova, N., Stoian, R., Rosenfeld, A., Hertel, I., Marine, W., and Campbell, E. *Applied Physics A* **81**, 345–356 (2005).

[25] Bonse, J. and Krüger, J. *Journal of Applied Physics* **108**, 034903 (2010).

[26] Derrien, T. J.-Y., Sarnet, T., Sentis, M., and Itina, T. E. *Journal of Optoelectronics and Advanced Materials* **12**(3), 610–615 March (2010).

**Figure captions**

Figure 1 (a) - SEM picture showing the effect of a single laser pulse with a homogenized profile, 1.15 J/cm² fluence, 100 fs of pulse duration, and 800 nm wavelength. Concentric modifications appear around defects. Period is near laser wavelength. The inset corresponds to the simulated zone.

Figure 1(b) – The average laser power deposited on a silicon surface by laser irradiation with 100 fs laser pulse (flat Si with modeled defects). Result of 3D FDTD simulation performed by using an absorptive layer at the boundaries.

Figure 2. Period of a surface plasmon polariton wave as a function of electron density for different collision frequencies. The critical electron density is $1.73 \; 10^{27}$ m$^{-3}$.

Figure 3. Conditions on free electron density obtained by using the Drude model to initiate plasmon resonance. Green dashed curve shows the free-electron density needed to reach the metallic condition (as a function of collision frequency). The red plain curve is the free-electron density necessary to excite a surface plasmon polariton.

Figure 4. (i) – The critical angle (green dashed curve) and SPP resonance angle (red plain curve) as a function of free electron density at vacuum/silicon interface for a given collision frequency. (ii) Red plain curve is for the free electron density permitting SPP phase-matching as a function of angle of incidence. Green dashed curve shows the maximum electron density, for which phase matching is possible.

Figure 5. Both calculated and experimentally measured periods of ripples normalized by laser wavelength as a function of laser fluence. The corresponding period is shown at the right axis. The spots obtained with equal number of pulses have been linked by splines interpolation

curves. The surface plasmon model has provided periods reported by a thick curve. Our single laser pulse experiments are shown by the cross-points with error bars on the top-right of the figure, and correspond to the SPP model. The results of Sipe's model are shown by using a thick curve with error bars measured on the simulated periods. Experimental points with low number (N > 1) of pulses are shown by star and square points Note that several points can exist for a same fluence.

(a) 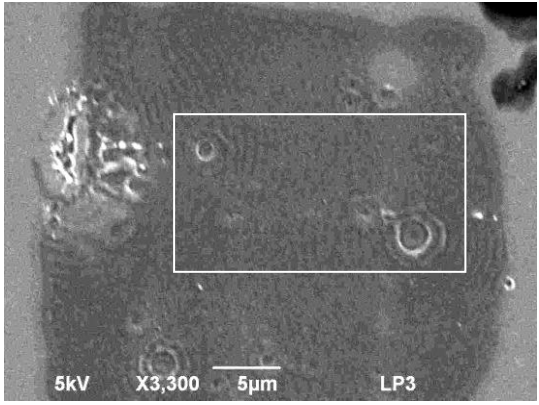

(b) 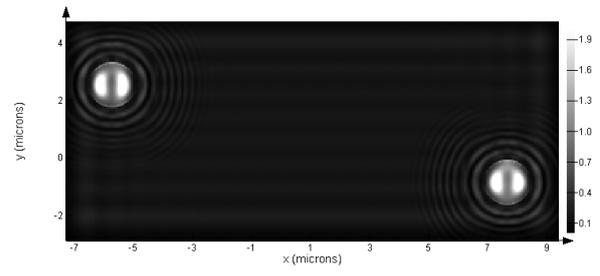

**Figure 1**

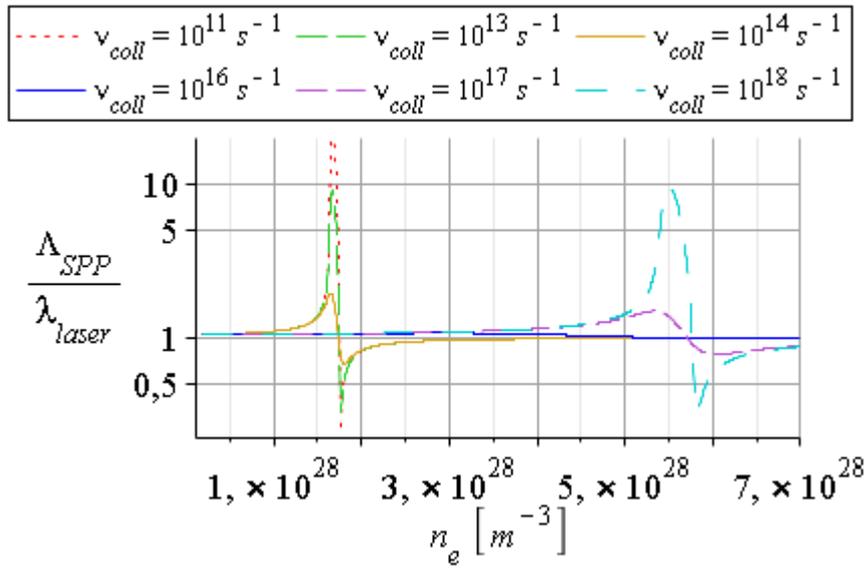

Figure 2

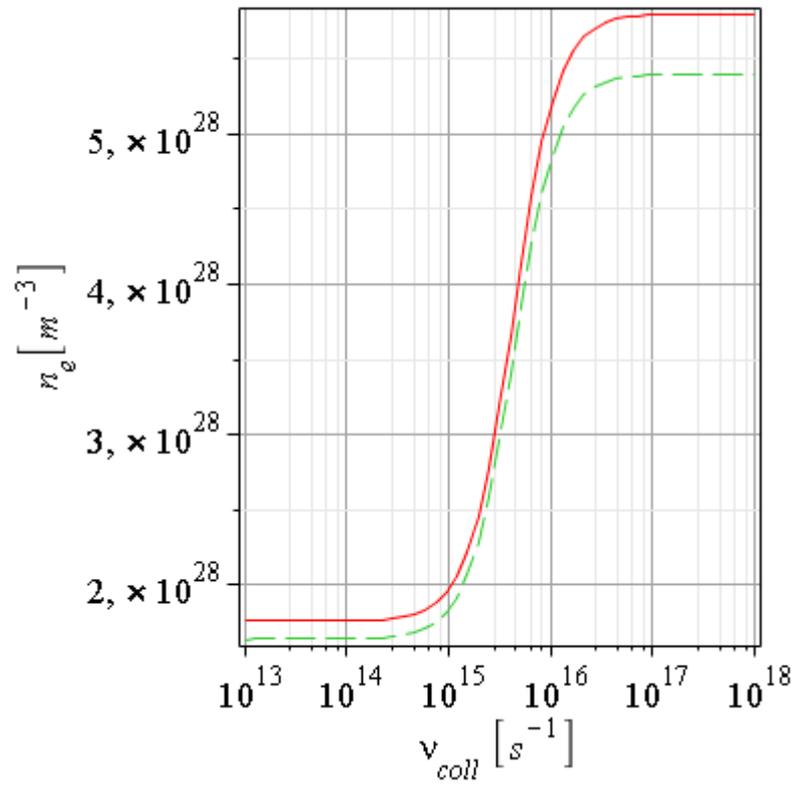

**Figure 3.**

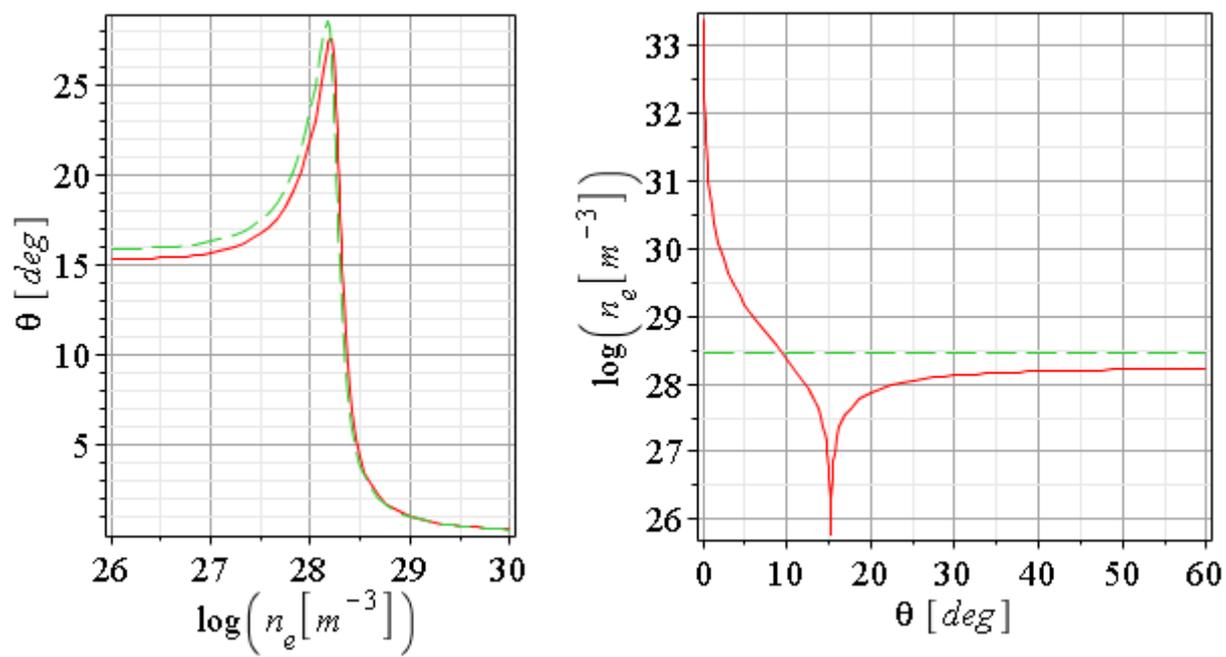

**Figure 4.**

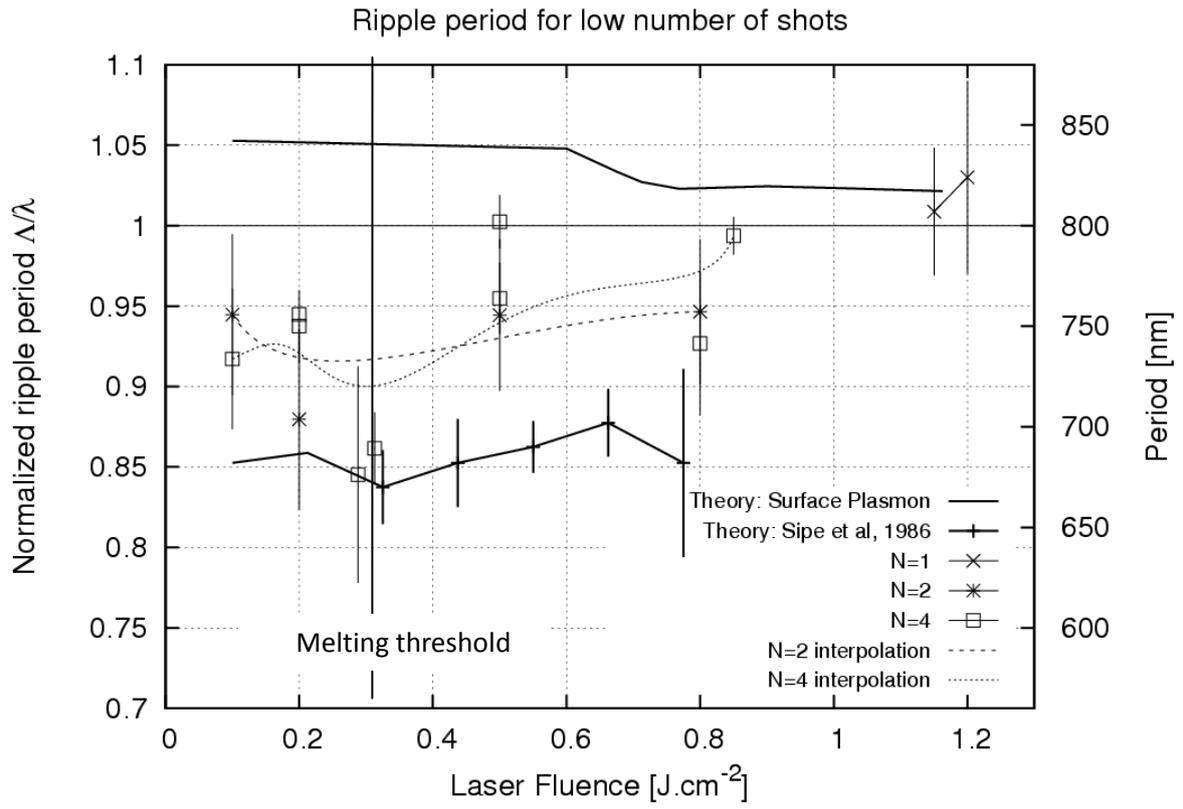

Figure 5.